\documentclass{Rinton-P9x6}

\def\ga{\mathrel{\raise.3ex\hbox{$>$\kern-.75em\lower1ex\hbox{$\sim$}}}}
\def\la{\mathrel{\raise.3ex\hbox{$<$\kern-.75em\lower1ex\hbox{$\sim$}}}}
\def\beq{\begin{equation}}  
\def\eeq{\end{equation}}
\def\PL{{\it Phys.Lett.} }
\def\PR{{\it Phys.Rev.} }

\def\NP{{\it Nucl. Phys.} }
\def\ohsq{\Omega_{\chi} h^2}

\def\m12{m_{1\!/2}}
\def\mst{m_{\tilde\tau_1}}

\def\st{{\widetilde \tau}_{\scriptscriptstyle\rm 1}}

\def\gev{{\rm \, Ge\kern-0.125em V}}
\def\ba{\begin{eqnarray}}
\def\ea{\end{eqnarray}}

\begin{document}

\title{Sugra Dark Matter\footnote{Summary of talk given at The International Conference 20 Years of
SUGRA and Search for SUSY and Unification (SUGRA20), Northeastern University, Boston MA,
March 2003.}}

\author{Keith A. Olive}

\address{William I. Fine Theoretical Physics
Institute, School of Physics and Astronomy, \\
University of Minnesota, Minneapolis, MN 55455 USA\\
E-mail:  olive@umn.edu}  


\maketitle

\abstracts{
\vskip -2.2in
\rightline{hep-ph/0308035}
\rightline{UMN--TH--2209/03}
\rightline{FTPI--MINN--03/20}
\rightline{August 2003}
\vskip 1.6in
I review the
phenomenological and cosmological constraints on the
parameter space associated with the Constrained Minimal Supersymmetric Standard Model (CMSSM).  The effect of the recent WMAP determination of the cold dark matter density will be discussed.  A very constrained model (based on minimal supergravity relations between the bi- and tri-linear supersymmetry breaking mass terms) will be outlined.  }

\section{Introduction}
It is well known that supersymmetric models with conserved $R$-parity
contain one new stable particle which is a candidate for cold dark matter
(CDM)\cite{EHNOS}. There are very strong constraints, however, forbidding the existence of stable or
long lived particles which are not color and electrically neutral.
Strong and electromagnetically interacting LSPs would become bound with
normal matter forming anomalously heavy isotopes. Indeed, there are very
strong upper limits on the abundances, relative to hydrogen, of nuclear
isotopes\cite{isotopes},
$n/n_H \la 10^{-15}~~{\rm to}~~10^{-29}
$
for 1 GeV $\la m \la$ 1 TeV. 
There are relatively few supersymmetric candidates which are not colored
and are electrically neutral.  The sneutrino\cite{snu,fos} is one
possibility, but in the MSSM, it has been excluded as a dark matter
candidate by direct\cite{dir} and indirect\cite{indir} searches.  Another possibility is the
gravitino which is probably the most difficult to exclude. 
I will concentrate on the remaining possibility in the MSSM, namely the
neutralinos.

 There are four neutralinos, each of which is a  
linear combination of the $R=-1$, neutral fermions\cite{EHNOS}: the wino
$\tilde W^3$, the partner of the
 3rd component of the $SU(2)_L$ gauge boson;
 the bino, $\tilde B$, the partner of the $U(1)_Y$ gauge boson;
 and the two neutral Higgsinos,  $\tilde H_1$ and $\tilde H_2$.
 In general,
neutralinos can  be expressed as a linear combination
\begin{equation}
	\chi = \alpha \tilde B + \beta \tilde W^3 + \gamma \tilde H_1 +
\delta
\tilde H_2
\end{equation}
The solution for the coefficients $\alpha, \beta, \gamma$ and $\delta$
for neutralinos that make up the LSP 
can be found by diagonalizing the mass matrix
\beq
      ({\tilde W}^3, {\tilde B}, {{\tilde H}^0}_1,{{\tilde H}^0}_2 )
  \left( \begin{array}{cccc}
M_2 & 0 & {-g_2 v_1 \over \sqrt{2}} &  {g_2 v_2 \over \sqrt{2}} \\
0 & M_1 & {g_1 v_1 \over \sqrt{2}} & {-g_1 v_2 \over \sqrt{2}} \\
{-g_2 v_1 \over \sqrt{2}} & {g_1 v_1 \over \sqrt{2}} & 0 & -\mu \\
{g_2 v_2 \over \sqrt{2}} & {-g_1 v_2 \over \sqrt{2}} & -\mu & 0 
\end{array} \right) \left( \begin{array}{c} {\tilde W}^3 \\
{\tilde B} \\ {{\tilde H}^0}_1 \\ {{\tilde H}^0}_2 \end{array} \right)
\eeq
where $M_1 (M_2)$ are the soft supersymmetry breaking
 U(1) (SU(2))  gaugino mass terms. $\mu$ is the supersymmetric Higgs
mixing mass parameter and since there are two Higgs doublets in
the MSSM, there are  two vacuum expectation values, $v_1$ and $v_2$. One
combination of these is related to the $Z$ mass, and therefore is not a
free parameter, while the other combination, the ratio of the two vevs,
$\tan \beta$, is free.

The most general version of the MSSM, despite its minimality in particles and
interactions contains well over a hundred new parameters. The study of such a model
would be untenable were it not for some (well motivated) assumptions.
These have to do with the parameters associated with supersymmetry breaking.
It is often assumed that, at some unification scale, all of the gaugino masses
receive a common mass, $m_{1/2}$. The gaugino masses at the weak scale are
determined by running a set of renormalization group equations.
Similarly, one often assumes that all scalars receive a common mass, $m_0$,
at the GUT scale (though one may wish to make an exception for the Higgs
soft masses). These too are run down to the weak scale. The remaining
supersymmetry breaking parameters are the trilinear mass terms, $A_0$,
which I will also assume are unified at the GUT scale,  and the bilinear
mass term
$B$. There are, in addition, two physical CP violating phases which will
not be considered here.

The natural boundary conditions at the GUT scale for the MSSM would
include
$\mu$, the two soft Higgs masses ($m_1$ and $m_2$) and $B$ in addition to
$m_{1/2}$,
$m_0$, and $A_0$. In this case, upon running the RGEs down to a low energy
scale and minimizing the Higgs potential, one would predict the values of $M_Z$, 
$\tan \beta$, and the Higgs
pseudoscalar mass, $m_A$ (in addition to all of the sparticle masses).
Since $M_Z$ is known, it is more useful to analyze supersymmetric models
where $M_Z$ is input rather than output.  It is also common to treat
$\tan \beta$ as an input parameter. This can be done at the expense of 
shifting $\mu$ (up to a sign) and $B$ from inputs to outputs. 
When the supersymmetry breaking Higgs soft masses are also unified at the
GUT scale (and take the common value $m_0$), the model is often referred
to as the constrained MSSM or CMSSM. Once these parameters are set, the
entire spectrum of sparticle masses at the weak scale can be calculated. 

\section{Constraints}

\subsection{The Relic Density}

The relic abundance of LSP's is 
determined by solving
the Boltzmann
 equation for the LSP number density in an expanding Universe.
 The technique\cite{wso} used is similar to that for computing
 the relic abundance of massive neutrinos\cite{lw}.
The relic density depends on additional parameters in the MSSM beyond $m_{1/2},
\mu$, and $\tan \beta$. These include the sfermion masses, $m_{\tilde
f}$, as well as  $m_A$, all derived from $m_0$, $A_0$, and
$m_{1/2}$. To determine the relic density it is necessary to obtain the
general annihilation cross-section for neutralinos.  In much of the
parameter space of interest, the LSP is a bino and the annihilation
proceeds mainly through sfermion exchange. Because of the p-wave
suppression associated with Majorana fermions, the s-wave part of the
annihilation cross-section is suppressed by the outgoing fermion masses. 
This means that it is necessary to expand the cross-section to include
p-wave corrections which can be expressed as a term proportional to the
temperature if neutralinos are in equilibrium. 
The partial wave expansion results in a very good approximation to the relic density
expect near s-channel annihilation poles,  thresholds and in regions where the LSP is
nearly degenerate with the next lightest supersymmetric particle\cite{gs}, where a 
more accurate treatment is necessary.

The preferred range of the relic LSP density has always been restricted to a relatively narrow range $0.1 < \Omega_{\rm CDM} h^2 < 0.3$, where values much smaller than
the lower bound are disfavoured by analyses
of structure formation in the CDM framework and the upper limit assumes only that the age of the Universe is $> 12$ Gyr.  However, one should note that the LSP
may not constitute all the CDM, in which case
$\Omega_{\rm LSP}$ could be reduced below this value.
This range has, however, been altered
significantly by the recent improved determination of the allowable range
of the cold dark matter density obtained by combining WMAP and other
cosmological data:  $0.094 < \Omega_{CDM} < 0.129$ at the 2-$\sigma$
level\cite{WMAP}.

\subsection{Accelerator Bounds}

The most relevant constraints on the supersymmetric parameter
space are: $m_{\chi^\pm} > 104$~GeV \cite{LEPsusy}, $m_{\tilde e} > 99$~GeV 
 \cite{LEPSUSYWG_0101} and $m_h >
114$~GeV \cite{LEPHiggs}. The former two constrain $m_{1/2}$ and $m_0$ directly
via the sparticle masses, and the latter indirectly via the sensitivity of
radiative corrections to the Higgs mass to the sparticle masses,
principally $m_{\tilde t, \tilde b}$. The  latest version
of {\tt FeynHiggs}~\cite{FeynHiggs} is used for the calculation of $m_h$. 
The Higgs limit  imposes important constraints
principally on $m_{1/2}$ particularly at low $\tan \beta$.
Another constraint is the requirement that
the branching ratio for $b \rightarrow
s \gamma$ is consistent with the experimental measurements\cite{bsg}. 
These measurements agree with the Standard Model, and
therefore provide bounds on MSSM particles,  such as the chargino and
charged Higgs masses, in particular. Typically, the $b\rightarrow s\gamma$
constraint is more important for $\mu < 0$, but it is also relevant for
$\mu > 0$,  particularly when $\tan\beta$ is large. The constraint imposed by
measurements of $b\rightarrow s\gamma$ also excludes small
values of $m_{1/2}$. 

Finally, there are
regions of the $(m_{1/2}, m_0)$ plane that are favoured by
the BNL measurement\cite{newBNL} of $g_\mu - 2$ at the 2-$\sigma$ level, corresponding to 
a deviation of $(33.9 \pm 11.2) \times 10^{-10}$ from the Standard Model 
calculation\cite{Davier} using $e^+ e^-$ data.  One should be  however 
aware that this constraint is still under discussion and it will not be used to
constrain $\tan \beta$. All the $\mu > 0$
planes would be consistent with $g_\mu - 2$ at the 3-$\sigma$ level,
whereas $\mu < 0$ is disfavoured even if one takes a relaxed view of the
$g_\mu - 2$ constraint.

\section{Results for the $m_{1/2} - m_0$ Plane Before and After WMAP}

\begin{figure}[h]
\begin{minipage}{8in}
\hspace*{0.3in}
\epsfig{file=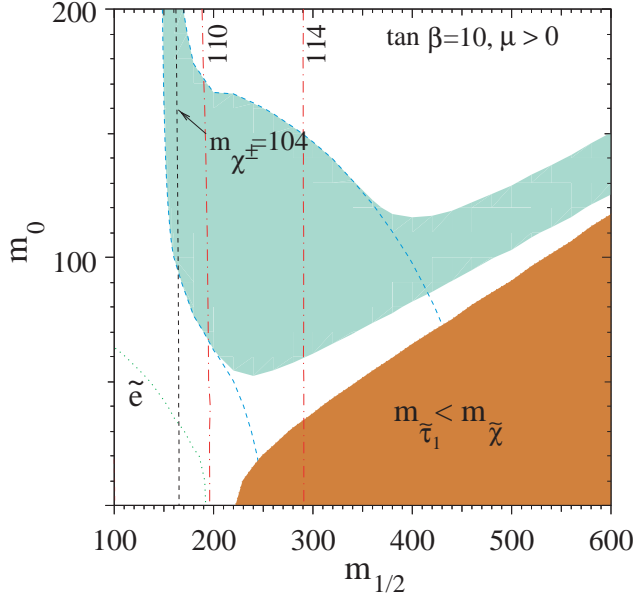,height=3.2in} \hfill
\end{minipage}
\caption{\label{fig:sm}
{\it The light-shaded `bulk' area is the cosmologically preferred 
region with \protect\mbox{$0.1\leq\ohsq\leq 0.3$}.   The light dashed lines
show the location  of the cosmologically preferred region  if one
ignores  coannihilations with the light sleptons.  
In the dark shaded region in the bottom right, the LSP is
the ${\tilde
\tau}_1$, leading to an unacceptable abundance
of charged dark matter.  Also shown is the isomass
contour $m_{\chi^\pm} = 104$~GeV and $m_h = 110,114$~GeV,
as well as an indication of the slepton bound from
LEP.  }}
\end{figure}

As noter earlier,  the spectrum for a given model in the CMSSM is determined by
the parameter set $m_{1/2}, m_0, \tan \beta,  A_0$, and the sign of $\mu$. For now,  I will assume that $A_0 = 0$.  Then, for a given value of $\tan \beta$ and $sgn(\mu)$, the resulting constraints
can be displayed on the  $m_{1/2} - m_0$ plane.
Let us first consider the region of parameter space with relatively low values of 
$m_{1/2}$ and $m_0$, i.e. the  `bulk' region in the CMSSM for $\tan \beta =
10$ and $\mu > 0$ shown in Fig. \ref{fig:sm} \cite{efo}. 
 The light
shaded region correspond to
\mbox{$0.1<\ohsq<0.3$}.  The dark shaded region has $m_{\st}< m_\chi$
and is excluded. The light dashed contours
indicate the corresponding region in $\ohsq$ if one ignores the effect of
coannihilations.  Neglecting coannihilations, one would find an upper
bound of $\sim450\gev$ on $\m12$, corresponding to an upper bound of
roughly $200\gev$ on $m_{\tilde B}$.

Coannihilations are important when there are several
particle species $i$, with different masses, and
each with its own number density $n_i$ and 
equilibrium number density $n_{0,i}$.
In this case inclusion of coannihilations is required because of  the degeneracy of $\chi$ and $\st$.
The effect of coannihilations is
to create an allowed band about 25-50 $\gev$ wide in $m_0$ for $\m12 \la
1400\gev$, which tracks above the $\mst=m_\chi$ contour\cite{efo,moreco}.

\begin{figure}[h]
\begin{minipage}{8in}
\epsfig{file=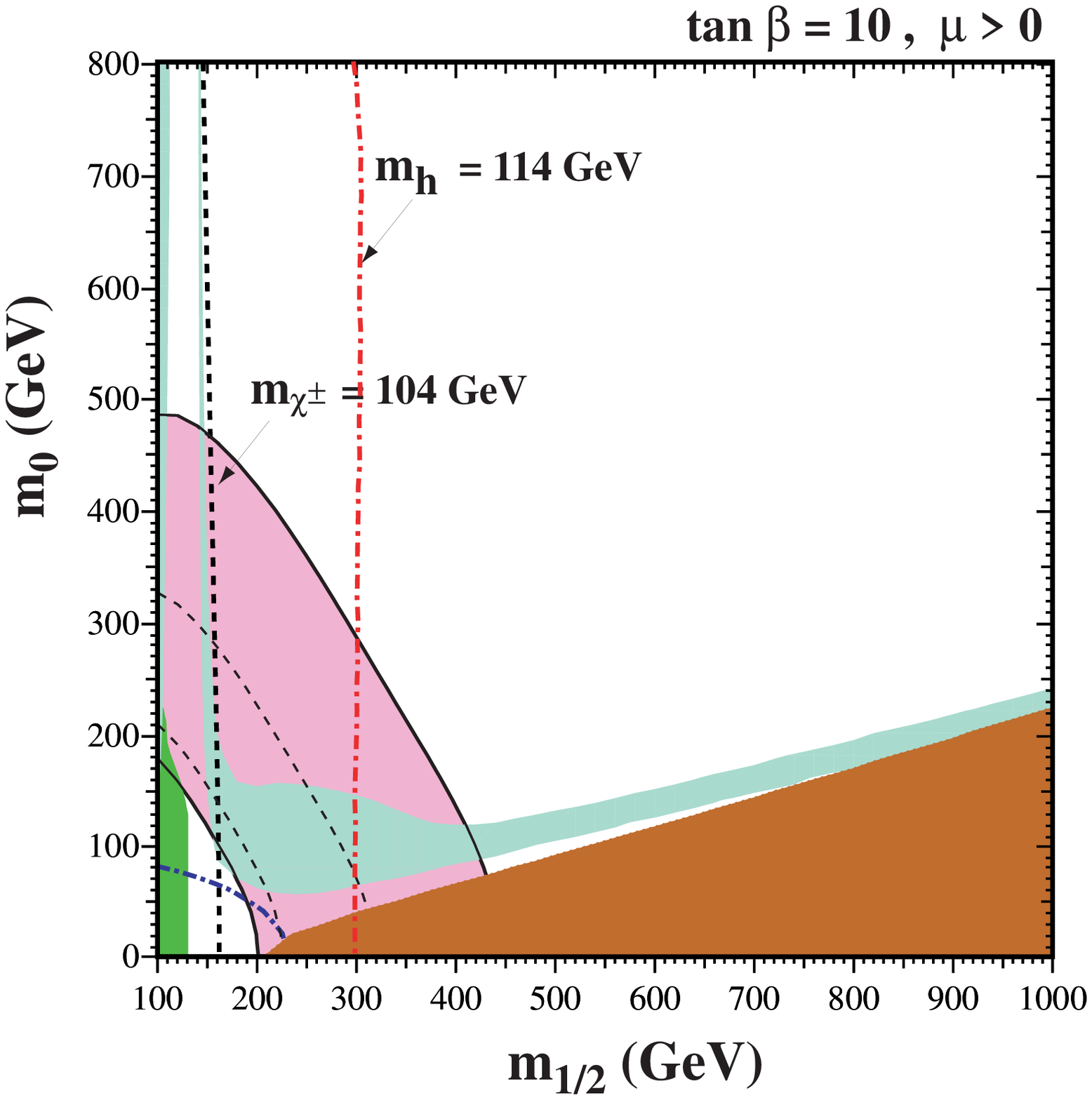,height=2.4in}
\epsfig{file=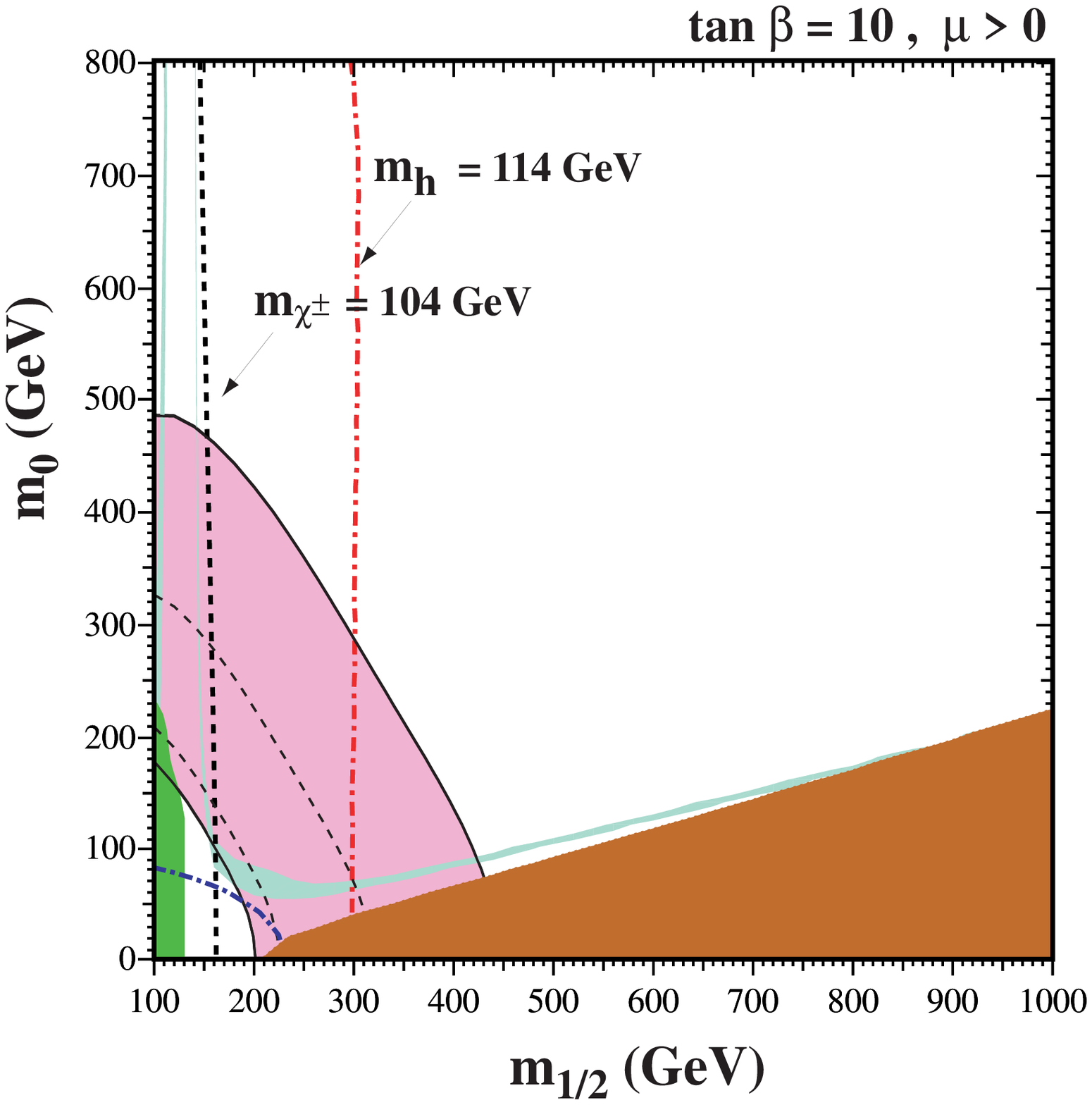,height=2.4in} \hfill
\end{minipage}
\caption{\label{fig:UHM}
{\it The $(m_{1/2}, m_0)$ planes for  (a) $\tan \beta = 10$ and  $\mu > 0$, 
assuming $A_0 = 0, m_t = 175$~GeV and
$m_b(m_b)^{\overline {MS}}_{SM} = 4.25$~GeV. The near-vertical (red)
dot-dashed lines are the contours $m_h = 114$~GeV as calculated using
{\tt FeynHiggs}~\protect\cite{FeynHiggs}, and the near-vertical (black) dashed
line is the contour $m_{\chi^\pm} = 104$~GeV. Also
shown by the dot-dashed curve in the lower left is the corner
excluded by the LEP bound of $m_{\tilde e} > 99$ GeV. The medium (dark
green) shaded region is excluded by $b \to s
\gamma$, and the light (turquoise) shaded area is the cosmologically
preferred regions with \protect\mbox{$0.1\leq\ohsq\leq 0.3$}. In the dark
(brick red) shaded region, the LSP is the charged ${\tilde \tau}_1$. The
region allowed by the E821 measurement of $a_\mu$ at the 2-$\sigma$
level, is shaded (pink) and bounded by solid black lines, with dashed
lines indicating the 1-$\sigma$ ranges. In (b), the relic density is restricted to the
range $0.094 < \ohsq < 0.129$. }}
\end{figure}

A larger view of the $\tan \beta = 10$ parameter plane is shown in the left panel of Fig.
\ref{fig:UHM} \cite{efo,EFGO,EFGOSi,eos3}.  Included here are
the phenomenological constraints discussed above.  
As one can see, the region preferred by $g-2$
overlaps very nicely with the `bulk' region for $\tan \beta = 10$ and
$\mu > 0$. In the second panel of Fig.
\ref{fig:UHM}, we see the effect of imposing the WMAP range on the
neutralino density\cite{eoss,morewmap}.
We see immediately that (i) the cosmological regions are
generally much narrower, and (ii) the `bulk' regions at small $m_{1/2}$
and $m_0$ have almost disappeared, in particular when the laboratory
constraints are imposed. Looking more closely at the coannihilation
regions, we see that (iii) they are significantly truncated as well as
becoming much narrower, since the reduced upper bound on $\Omega_\chi h^2$
moves the tip where $m_\chi = m_{\tilde \tau}$ to smaller $m_{1/2}$
so that the upper limit is now $m_{1/2} \la 950$ GeV or $m_\chi \la 400$ GeV. 

\begin{figure}[hbtp]
\begin{minipage}{8in}
\epsfig{file=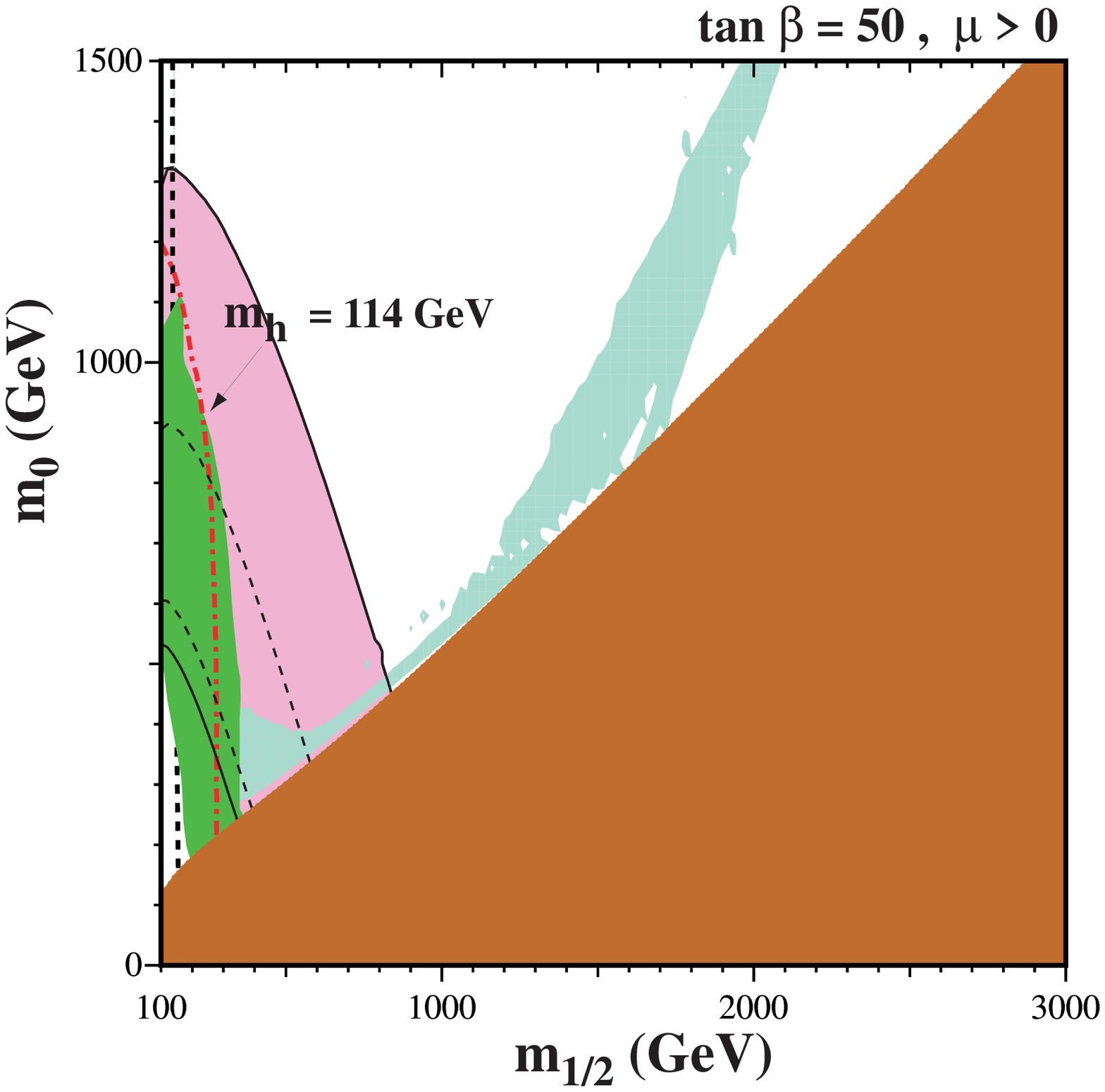,height=2.3in}
\epsfig{file=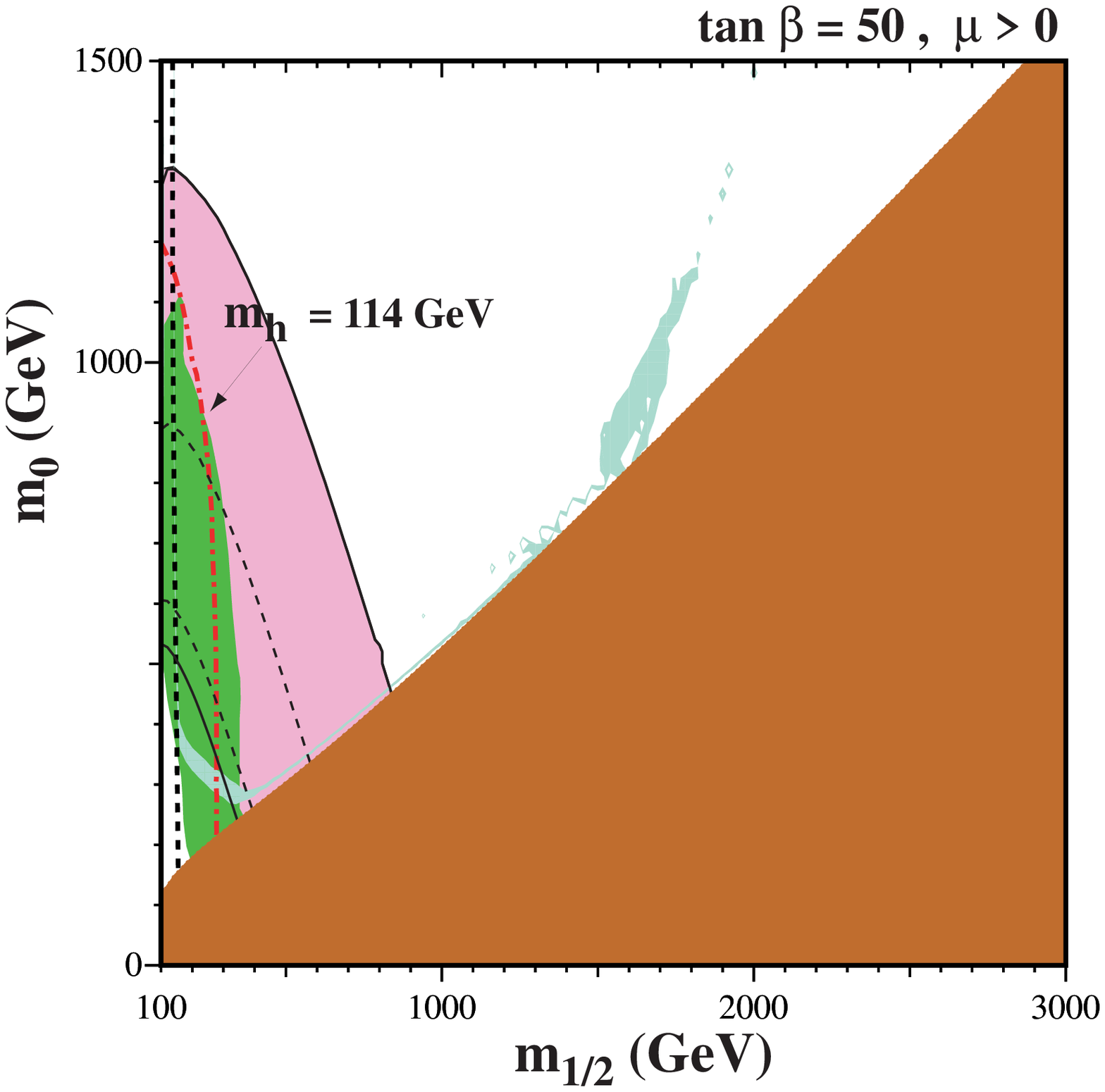,height=2.3in}
\end{minipage}
	\caption{\it As in Fig. \protect\ref{fig:UHM} for  $\tan \beta = 50$ and $\mu > 0$ }
	\label{rd2c50}
\end{figure}

Another
mechanism for extending the allowed CMSSM region to large
$m_\chi$ is rapid annihilation via a direct-channel pole when $m_\chi
\sim {1\over 2} m_{A}$~\cite{funnel,EFGOSi}. Since the heavy scalar and
pseudoscalar Higgs masses decrease as  
$\tan \beta$ increases, eventually  $ 2 m_\chi \simeq  m_A$ yielding a
`funnel' extending to large
$m_{1/2}$ and
$m_0$ at large
$\tan\beta$, as seen in Fig.~\ref{rd2c50}.
The difficulty and necessary care involved in calculations at large 
$\tan \beta$ were previously discussed\cite{EFGOSi}. For related CMSSM
calculations see \cite{otherOmega}. 

In the second panel of Fig.
\ref{rd2c50}, we see the effect of imposing the WMAP range on the
neutralino density\cite{eoss}. We see rapid-annihilation funnels that
are also narrower and extend to lower $m_{1/2}$ and $m_0$ than
previously. They weaken significantly the upper bound on $m_\chi$ for
$\tan \beta \ga 50$ for $\mu
> 0$.

\begin{figure}
\begin{center}
\mbox{\epsfig{file=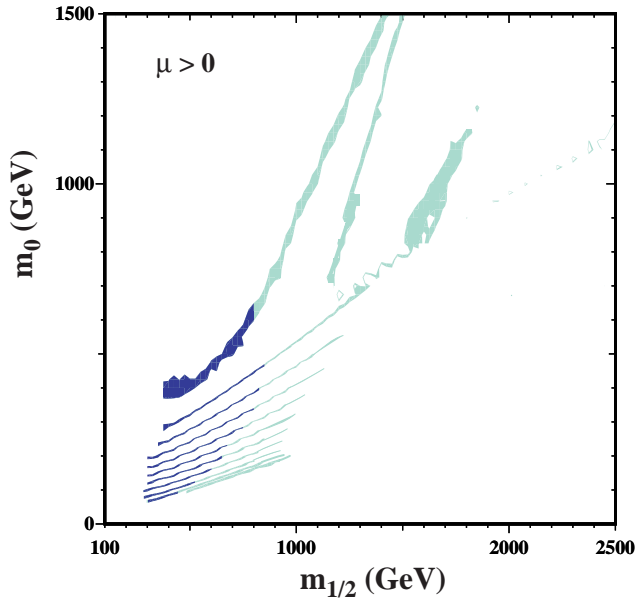,height=3.2in}}
\end{center}
\caption{\label{fig:strips}\it
The strips display the regions of the $(m_{1/2}, m_0)$ plane that are
compatible with $0.094 < \Omega_\chi h^2 < 0.129$ and the laboratory
constraints for $\mu > 0$ and $\tan \beta = 5, 10, 15, 20, 25, 30,
35, 40, 45, 50, 55$. The parts of the strips compatible with $g_\mu - 2$ 
at the 2-$\sigma$ level have darker shading.
}
\end{figure}

Shown in Fig.~\ref{fig:strips} are the wmap lines\cite{eoss} of the $(m_{1/2}, m_0)$
plane allowed by the new cosmological constraint $0.094 < \Omega_\chi h^2
< 0.129$ and the laboratory constraints listed above, for $\mu > 0$ and
values of $\tan \beta$ from 5 to 55, in steps $\Delta ( \tan \beta ) = 5$.
We notice immediately that the strips are considerably narrower than the
spacing between them, though any intermediate point in the $(m_{1/2},
m_0)$ plane would be compatible with some intermediate value of $\tan
\beta$. The right (left) ends of the strips correspond to the maximal
(minimal) allowed values of $m_{1/2}$ and hence $m_\chi$. 
The lower bounds on $m_{1/2}$ are due to the Higgs 
mass constraint for $\tan \beta \le 23$, but are determined by the $b \to 
s \gamma$ constraint for higher values of $\tan \beta$. The
upper bound on $m_{1/2}$ for $\tan \beta
\ga 50$ is clearly weaker, because of the rapid-annihilation regions.

\section{Using SUGRA to constrain the CMSSM}

While the CMSSM models described above are certainly
mSUGRA inspired, minimal supergravity models
can be argued to be still more predictive or constrained.  
Let us begin by assuming that supersymmetry is broken in a hidden sector
so that the superpotential can be written as a sum of two terms, $W = F(\phi) +g(\zeta)$,
where $\phi$ represents all observable fields and $\zeta$ all hidden sector fields.
Then the scalar potential in an $N=1$ supergravity theory with minimal K\"ahler potential
$K(\phi,\zeta) = \phi \phi^* + \zeta \zeta^* + \ln |W|^2$, can be written as\cite{sugr2}
\ba
V & = & e^{(|\zeta|^2 + |\phi|^2)} \left[ |{\partial g \over \partial 
\zeta} + \zeta^* (g(\zeta) + F(\phi) )|^2 \right. \nonumber \\
& & + \left. |{\partial F \over \partial \phi} +
\phi^* (g(\zeta) + F(\phi) )|^2 - 3 | g(\zeta) + F(\phi) |^2 \right].
\label{cpot0}
\ea
We furthermore must choose $g(\zeta)$ such that when $\zeta$ picks up a vacuum
expectation value, supersymmetry is broken.
When the potential is expanded and terms inversely proportional to Planck mass are dropped,
one finds\cite{BIM,mark} 1) scalar mass universality with $m_0 = \langle g \rangle$; 
2) trilinear mass universality with 
$A_0 = \langle dg/d\zeta \rangle \langle \zeta \rangle + \langle g \rangle \langle \zeta \rangle^2$; 
and 3) $B_0 = A_0 - m_0$.

In fact, one of the primary motivations for the CMSSM, and for scalar mass
universality in particular, comes from the simplest model for local
supersymmetry breaking\cite{pol}, which involves just one additional
chiral multiplet $\zeta$ with a superpotential of the form
\beq
g(\zeta) \; = \; \nu(\zeta + \beta)
\label{polonyi}
\eeq
with $\vert \beta \vert = 2 - \sqrt{3}$, ensuring that the cosmological constant $\Lambda = 0$. 
After inserting the vev for
$\zeta$, $\langle \zeta \rangle = \sqrt{3} - 1$. The 
scalar potential in this model takes the form\cite{bfs}: 
\ba
V & = & e^{(4 - 2\sqrt{3})} \left[ |\nu + (\sqrt{3} -
1) (\nu + F(\phi)) |^2 \right. \nonumber \\
& & \left. +|{\partial F \over
\partial \phi} + \phi^* (\nu + F(\phi) )|^2 - 3 | \nu + F(\phi) |^2
\right] \nonumber \\
& = & e^{(4 - 2\sqrt{3})} |{\partial F \over \partial
\phi}|^2 \nonumber \\
& & + m_{3/2} e^{(2 - \sqrt{3})}(\phi {\partial F
\over \partial \phi} - \sqrt{3} F + h.c.) )  + m_{3/2}^2 \phi
\phi^* ,
\label{cpot}
\ea
where the gravitino mass is given by $m_{3/2} = e^{2-\sqrt{3}} \nu$.

First, up to an overall rescaling of the superpotential, $F \to
e^{\sqrt{3}-2} F$, the first term is the ordinary $F$-term part of the
scalar potential of global supersymmetry. The next term, which is
proportional to $m_{3/2}$, provides universal trilinear soft
supersymmetry-breaking terms $A = (3 -
\sqrt{3}) m_{3/2}$ and bilinear
soft supersymmetry-breaking terms $B = (2 - \sqrt{3}) m_{3/2}$, i.e., a
special case of the general relation  above between $B$ and
$A$. Finally, the last term represents a universal scalar mass of the type
advocated in the CMSSM, with $m_0^2 = m_{3/2}^2$, since the cosmological
constant $\Lambda$ vanishes in this model, by construction.

Given a relation between $B_0$ and $A_0$, we can no longer use the
standard CMSSM boundary conditions, in which $m_{1/2}$, $m_0$, $A_0$,
$\tan \beta$, and $sgn(\mu)$ are input at the GUT scale with $\mu$ and $B$ 
determined by the electroweak symmetry breaking condition.
Now, one is forced to input $B_0$ and instead $\tan \beta$ is 
calculated from the minimization of the Higgs potential\cite{eoss2}.
It should be stressed that this type of model is one that truly originates from minimal supergravity.

\begin{figure}
\begin{center}
\mbox{\epsfig{file=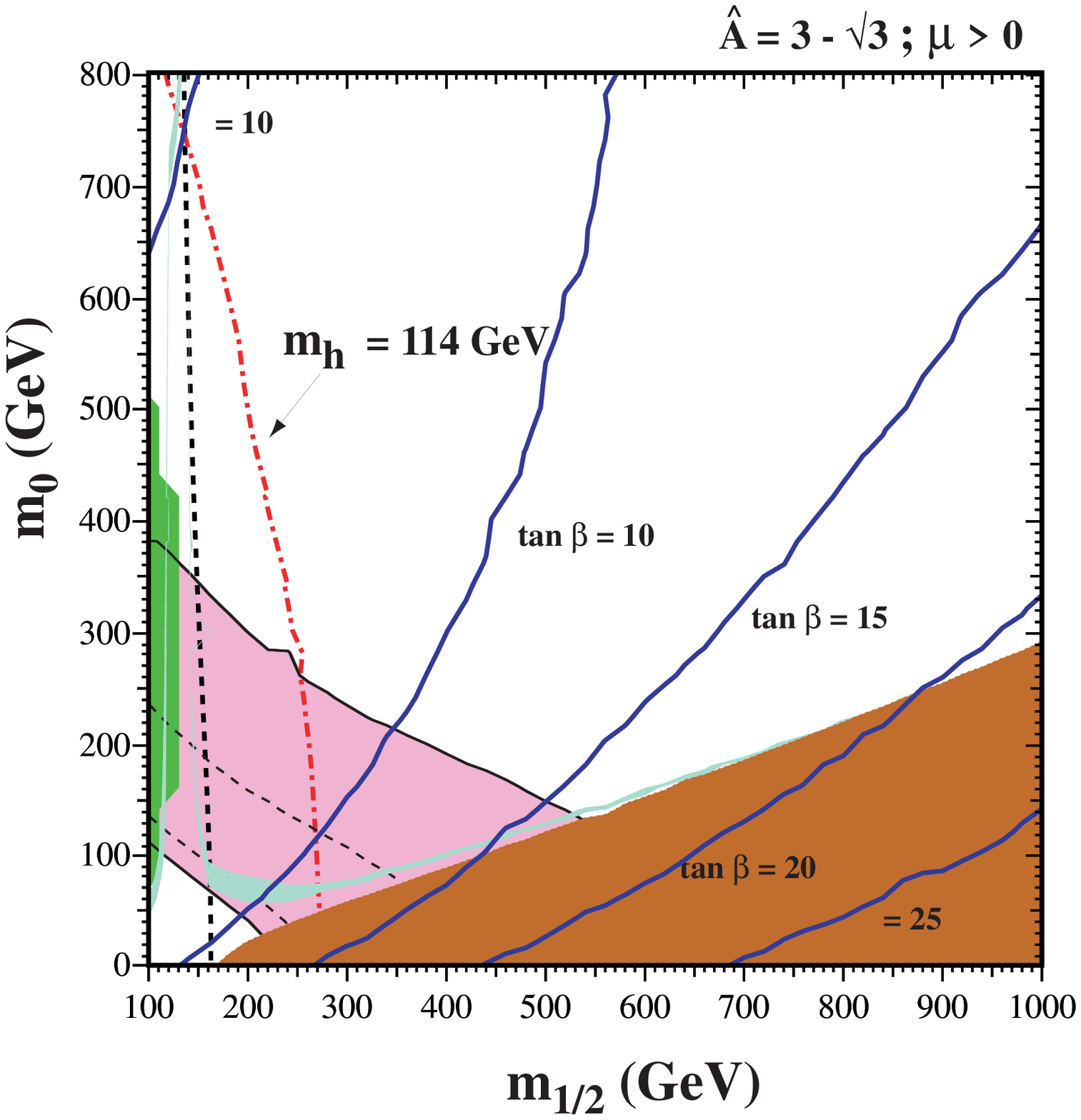,height=2.3in}}
\mbox{\epsfig{file=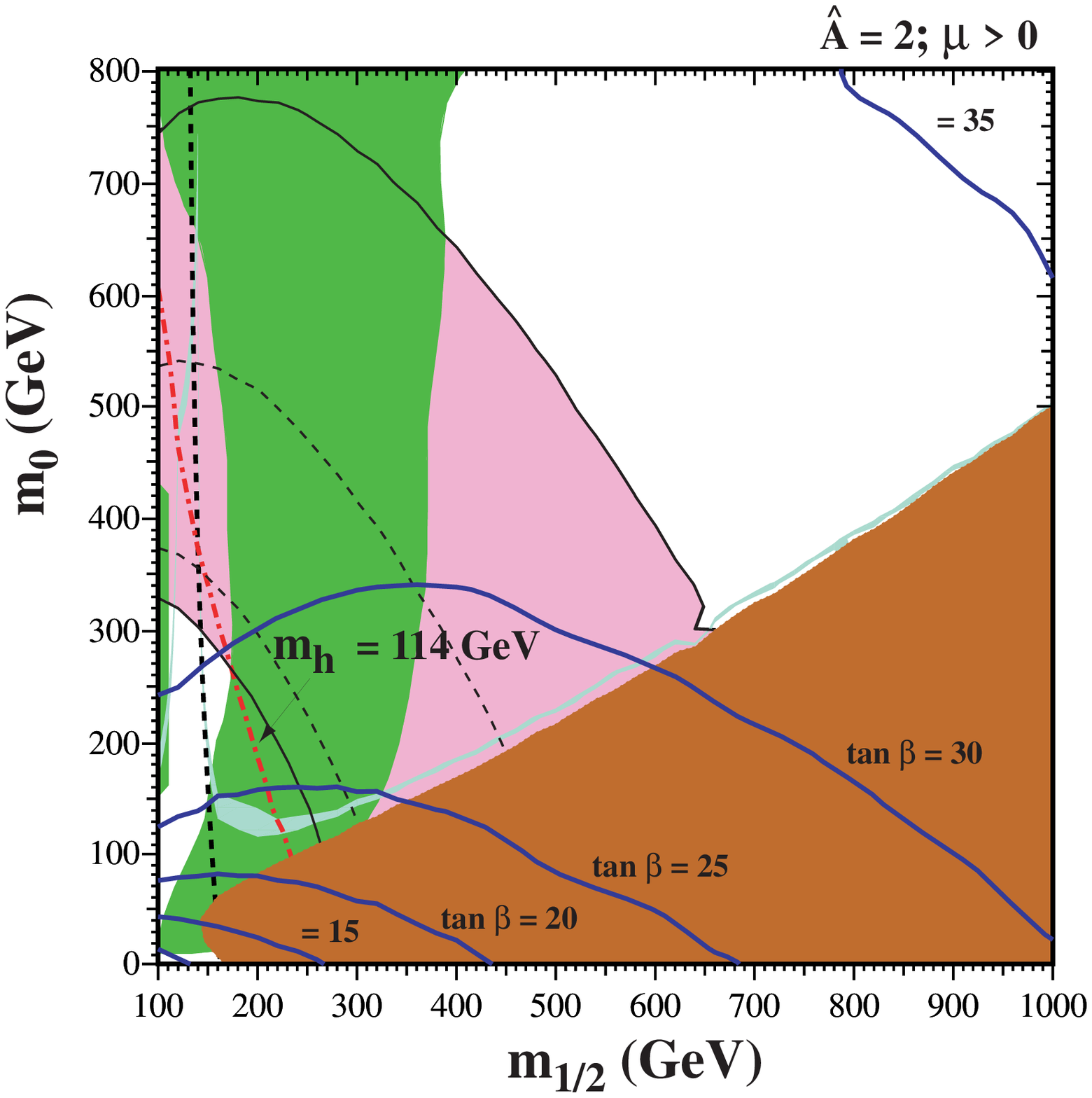,height=2.3in}}
\end{center}
\caption{\it
Examples of $(m_{1/2}, m_0)$ planes with contours of $\tan \beta$ 
superposed, for $\mu > 0$ and (a)  the simplest Polonyi model with ${\hat A} = 3 - 
\sqrt{3}, {\hat B} = {\hat A} -1$ and (b) ${\hat A} = 2.0, {\hat B} =
{\hat A} -1$. In each panel, we show the regions excluded by 
the LEP lower limits on MSSM particles, those ruled out by $b
\to s \gamma$ decay~\protect\cite{bsg} (medium green shading), and those 
excluded 
because the LSP would be charged (dark red shading). The region favoured 
by the WMAP range $\Omega_{CDM} h^2 =
0.1126^{+0.0081}_{-0.0091}$ has light turquoise shading. The region 
suggested by $g_\mu - 2$ is medium (pink) shaded.}
\label{fig:Polonyi}
\end{figure}

In Fig.~\ref{fig:Polonyi}, the contours of $\tan \beta$ (solid
blue lines) in the $(m_{1/2}, m_0)$ planes for two values of ${\hat
A}  = A_0/m_0$, ${\hat B} = B_0/m_0 = {\hat A} - 1$ and the sign of $\mu$ are displayed\cite{eoss2}. 
Also shown are the contours where
$m_{\chi^\pm} > 104$~GeV (near-vertical black dashed lines) and $m_h >
114$~GeV (diagonal red dash-dotted lines). The excluded regions where
$m_\chi > m_{\tilde \tau_1}$ have dark (red) shading, those excluded by $b
\to s \gamma$ have medium (green) shading, and those where the relic
density of neutralinos lies within the WMAP range $0.094 \le \Omega_\chi
h^2 \le 0.129$ have light (turquoise) shading. Finally, the regions
favoured by $g_\mu - 2$ at the 2-$\sigma$ level are medium (pink) shaded.

In panel (a) of Fig.~\ref{fig:Polonyi}, we
see that the Higgs constraint combined with the relic density requires $\tan \beta
\ga 11$, whilst the relic density also enforces $\tan \beta \la 
20$. For a given point in the $m_{1/2} - m_0$ plane, the calculated value of $\tan \beta$
increases as ${\hat A}$ increases.
This is seen in panel (b) of
Fig.~\ref{fig:Polonyi}, when ${\hat A} = 2.0$, close to its maximal value
for $\mu > 0$, the $\tan \beta$ contours turn over towards smaller
$m_{1/2}$, and only relatively large values $25 \la \tan \beta \la 35$ are
allowed by the $b \to s \gamma$ and $\Omega_{CDM} h^2$ constraints,
respectively.  For any given value of ${\hat A}$, there is only a relatively narrow
range allowed for $\tan \beta$.

At higher values of ${\hat A}$ (and high $\tan \beta)$, the off-diagonal 
elements  in the squark mass matrix become large at large $m_0$.
Therefore, we find no solutions which are phenomenologically viable
above  ${\hat A} \simeq 2.5$. This is
because the regions where the LSP is the ${\widetilde \tau}$ or the ${\widetilde t}$ close off
the parameter space. In fact, this feature is generic in the CMSSM as shown in
Fig. 3 of \cite{EFGO}. This effect is more severe at large $\tan\beta$, which 
further compounds the difficulty in going to large values of ${\hat A}$ in the type of models
discussed here. At very low values of ${\hat A}$ ($\la -1.9$), there are no viable solutions
due to the competition between the Higgs mass bound and the relic density.
At low ${\hat A}$, solutions give low values of $\tan \beta$ which in turn give low 
Higgs masses unless $m_{1/2}$ is very large.  For ${\hat A} < -1.9$, $m_{1/2}$ is pushed
passed the endpoint of the coannihilation region.

In addition, we note the absence of the funnel 
regions. This is due to the relatively small values of
$\tan \beta$ allowed in the class of models considered here: we recall
that the funnel region appears only for large $\tan \beta \ga 45$ for
$\mu > 0$ and $\tan \beta \ga 30$ for $\mu < 0$ in the CMSSM.

Rather than further constrain the CMSSM as described above, one can generalize the CMSSM case to include non-universal Higgs masses\cite{nonu,eos3}
(NUHM), in which case the 			
input parameters include $ \mu$ and $m_A,$ in addition to the standard CMSSM inputs.
In this case, the two soft Higgs masses,  $m_1,
m_2$ are no longer set equal to $m_0$ and are calculated from the electroweak symmetry breaking conditions. The NUHM parameter space was recently analyzed\cite{eos3} and a sample of the results
were reviewed in \cite{yudi}.

\section*{Acknowledgments}
This work was supported in part
by DOE grant DE--FG02--94ER--40823 at the University of Minnesota.

\end{document}